\documentclass[a4paper,11pt]{article}
\usepackage{jinstpub} 
\usepackage{lineno}
\usepackage{wrapfig}
\usepackage{siunitx}
\usepackage{svg}
\usepackage{subcaption}

\proceeding{Topical Workshop on Electronics for Particle Physics\\
  2025}

\title{MOSAIX Qualification System for ALICE ITS3}

\author{O. Groettvik}
\affiliation{European Organization for Nuclear Research (CERN),\\Geneva, Switzerland}
\collaboration[c]{on behalf of the ALICE Collaboration}

\emailAdd{ola.groettvik@cern.ch}

\abstract{The ALICE Inner Tracking System upgrade (ITS3) will employ stitched, wafer-scale Monolithic Active Pixel Sensors (MAPS) for the first time in high-energy physics. MOSAIX, a fully functional prototype and the final development step before production, measures 266 mm by 19 mm. The chip integrates 144 independently powered pixel matrices, eight 10~Gb/s transmitters, and on-chip power and data distribution.

This contribution presents the development of the MOSAIX test system and verification strategy, with emphasis on validating the testing infrastructure before chip availability. The system includes an FPGA that controls MOSAIX and parses its output, and a second MOSAIX emulator FPGA.
}

\begin{document}
\notoc
\maketitle
\flushbottom
\newpage
\clearpage
\setcounter{page}{1}

\section{Introduction}

The upgrade of the ALICE Inner Tracking System (ITS) for LHC Run 4, known as ITS3, aims to drastically reduce the material budget of the innermost tracking layers \cite{placeholder_its3_tdr}. This is achieved by constructing the detector layers from stitched, wafer-scale Monolithic Active Pixel Sensors (MAPS) that can be bent to form self-supporting structures. This design reduces the material budget to an unprecedented $0.09\,\%~X/X_0$ per layer, a factor of four improvement over its predecessor, ITS2.

The core component of this upgrade is MOSAIX, the final full-scale prototype sensor. Measuring \SI{266}{mm} long, MOSAIX is effectively a System-on-Chip (SoC) that integrates multiple standalone pixel sensor blocks with power distribution, control, and high-speed data aggregation directly onto the silicon. While this integration is key to the material budget reduction, it introduces challenges for testing and qualification. This paper presents the qualification system developed to validate the MOSAIX sensor. A central pillar of this effort is the strategy of achieving "day-one readiness," enabled by a purpose-built FPGA emulator of the sensor.

\section{The MOSAIX Sensor: A Qualification Challenge}

The complexity of MOSAIX stems from its high level of integration. The sensor comprises three main functional blocks stitched together: a Left Endcap (LEC), a Right Endcap (REC), and twelve Repeated Sensor Units (RSUs), all connected via a common stitched backbone.

Each of the 12 RSUs contains 12 independent sensor blocks, or "tiles," resulting in a total of 144 tiles on the MOSAIX segment. Each tile, comparable in functionality to the ALPIDE sensor~\cite{placeholder_alpide} used in the current ITS, includes its own pixel matrix, biasing unit, and digital periphery for control and readout. Furthermore, to explore analog frontend designs and biasing schemes, 12 distinct pixel design variants are implemented, with one variant in each of the 12 tiles of an RSU.

The system's complexity is further compounded by its services:
\begin{itemize}
    \item \textbf{Powering:} In addition to five global power domains, the chip features 288 local power domains for the individual analog and digital supplies of each tile. These are controlled by on-chip switches located in 48 distributed service nodes, requiring complex power-up sequences.
    \item \textbf{Data Transmission:} Data from the 144 tiles are transmitted on 144 parallel, clock-synchronous data links at \SI{160}{Mb/s} across distances of up to \SI{25}{cm} to the LEC. The LEC aggregates this data and transmits it off-detector via eight high-speed links, each operating at \SI{10}{Gb/s} and utilizing lpGBT encoding logic \cite{placeholder_lpgbt}.
\end{itemize}

The detector half-layers employ silicon sheets of 3 to 5 MOSAIX segments. To construct these layers one must dice these layers from wafers with 5 segments~\cite{placeholder_its3_ogrottvi}. One must thus have information about the yield and performance before the dicing.

\begin{figure}[h!]
\centering
\includegraphics[width=0.8\textwidth]{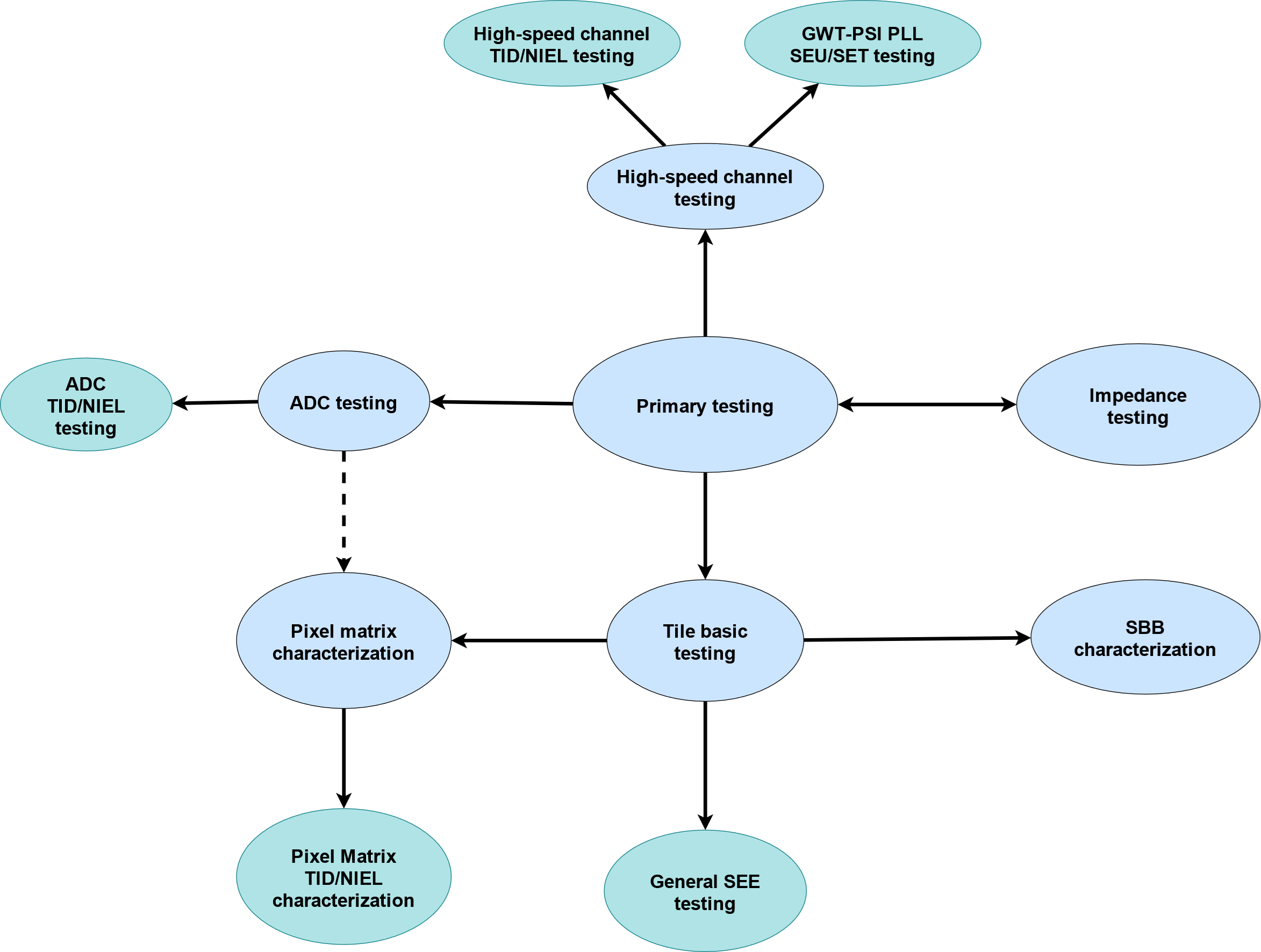}
\caption{A map of the major testing and characterization phases for MOSAIX. A substantial core of functionalities must be validated before pixel-level characterization can even begin, highlighting the system's high level of integration. ADC testing is a quasi-dependency for pixel matrix qualification, as it can be circumvented by using the external ADC path.}
\label{fig:dependency_map}
\end{figure}

The qualification task is therefore not merely the characterization of 144 independent sensors, but the validation of an entire integrated system where all components are interconnected on a single piece of silicon. The inability to pre-test individual components before integration demands a carefully planned qualification campaign that can navigate the intricate dependencies between the chip's subsystems. Figure~\ref{fig:dependency_map} shows the high-level test phases of the MOSAIX qualification campaign and their dependencies. Note that all major testing phases rely on a primary test phase, which comprises powering tests and ensuring that the full service layer of the MOSAIX is functioning, i.e. that all 48 service nodes of the MOSAIX are working properly. This phase is already a non-trivial test space.

\section{Qualification Strategy and System Architecture}

The primary objective of the qualification campaign is to provide timely feedback for the design of the final production sensor. The key goals include:
\begin{enumerate}
    \item Verification of all chip features and estimation of the production yield of critical blocks.
    \item Identification of at least one pixel variant that meets all performance specifications.
    \item Wafer-level qualification of the eight \SI{10}{Gb/s} high-speed links to inform the dicing strategy.
    \item Measurement of the radiation tolerance of newly designed components like the stitched backbone data buffers.
\end{enumerate}

\subsection{A Unified Test System}

To counterbalance the complexity of the MOSAIX, the qualification platform is based on a simple, standardized hardware setup. The core is an Intel Arria 10 FPGA system-on-module~\cite{enclustra_som} mounted on an Enclustra baseboard~\cite{enclustra_baseboard}, which provides control and data acquisition, while simple programmable laboratory power supplies are used for powering and monitoring.

\begin{figure}[h!]
\centering
\includegraphics[width=0.97\textwidth]{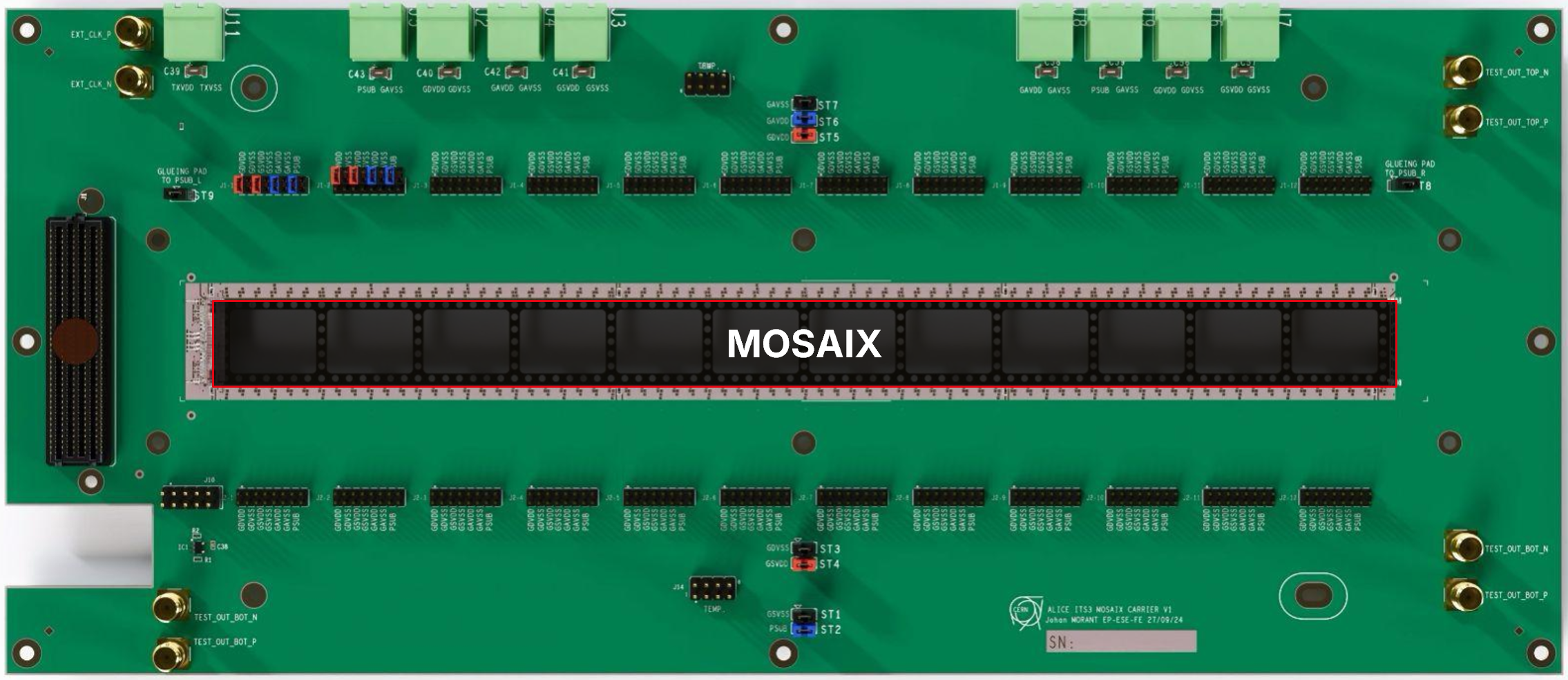}
\caption{\label{fig:carrier}The MOSAIX carrier card is one of the targets for the qualification system. It allows for deeper characterization and observability compared to testing on the wafer prober, e.g. the measurement of the IR-drop along the long edge (z-axis) of the sensor.}
\end{figure}

A key design principle is the "one system, many targets" approach. A standardized FMC connector on the readout board allows the same core system to interface with MOSAIX on different test targets carriers throughout the entire qualification lifecycle. This includes:
\begin{itemize}
    \item Wafer probing via a custom probe card.
    \item Characterization of single diced chips on a dedicated carrier board for detailed measurements, for example measuring the IR-drop along the z-axis of the sensor (see Figure~\ref{fig:carrier}).
    \item Commissioning of MOSAIX segments integrated in its final mechanical and electrical assembly in the Qualification Model of the ITS3 detector.
\end{itemize}
The system is also equipped to test external services like the external FPC. This approach minimizes redundant development effort and ensures consistency across all test stages.

\subsection{Control and Software Architecture}

The hardware is driven by a robust control architecture, achieved through modern firmware verification techniques and modular software intended for final use in the ALICE experiment. The FPGA firmware, which implements all interfaces to MOSAIX, has been rigorously verified using a UVM-based environment and continuous integration (CI) regression testing. The software architecture features a hardware-agnostic C++ API that abstracts away low-level FPGA and MOSAIX interactions. Furthermore, test sequence code is separated from the core API in a standalone repository, a modular design that simplifies community contributions and protects system-critical procedures from accidental changes.

\section{The MOSAIX Emulator}

\begin{figure}[h]
  \centering 
  \includegraphics[width=.97\textwidth]{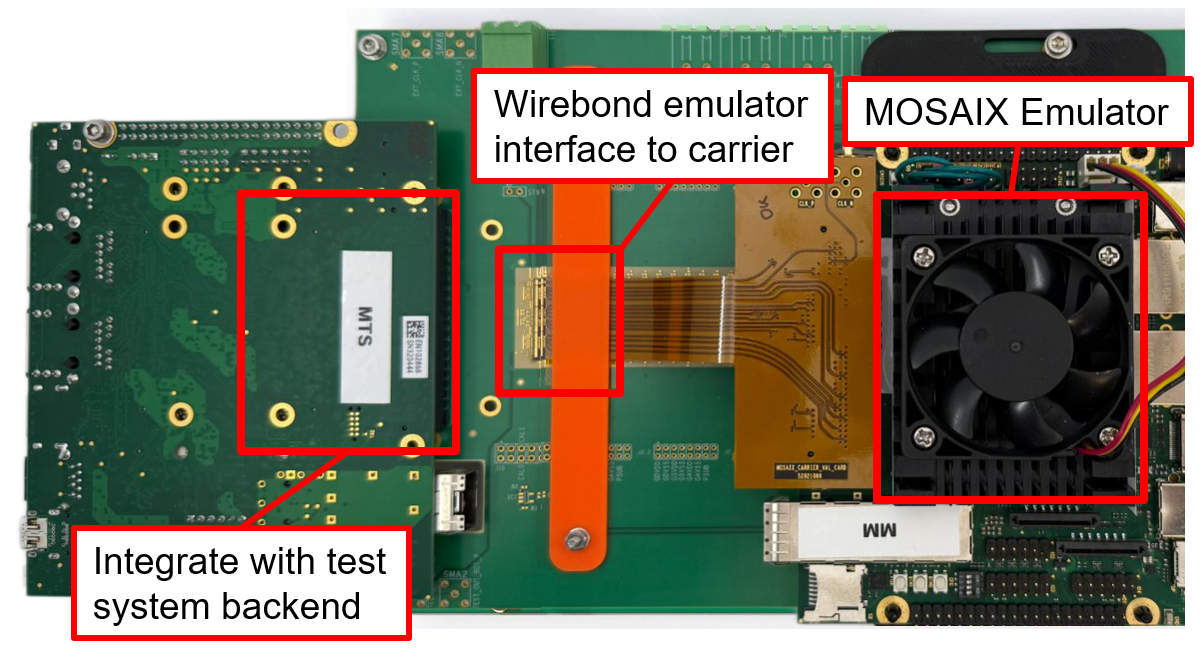}
  \caption{\label{fig:emulator_usage} Integration test setup for the MOSAIX segment carrier PCB.}
\end{figure}

A significant risk to the project timeline is the delayed discovery of integration issues, since comprehensive testing of the hardware, firmware, and software together requires the availability of the MOSAIX silicon. To enable parallel development and achieve "day-one readiness," a custom FPGA-based emulator of the MOSAIX sensor was developed. The MOSAIX emulator allows for integration tests of all the main system components. As an example, Figure~\ref{fig:emulator_usage} shows the integration test setup for the MOSAIX segment carrier card. The MOSAIX Emulator system is wire-bonded in place of the MOSAIX chip via a FPC. The qualification system backend board is connected to the carrier allowing applying stimuli and monitoring output. The functionality of the carrier PCB is qualified by employing normal control sequences of the qualification software, and reading out data over the 10 Gb/s links.

\subsection{Implementation}

The emulator was built on an identical Arria 10 FPGA platform as the test system, allowing for direct board-to-board connection. This setup perfectly replicates the physical hardware environment, including all eight \SI{10}{Gb/s} links, slow control, clock, and synchronization signals.

A full one-to-one model of MOSAIX is not feasible on the FPGA. Therefore, a balance was struck between model accuracy and available resources. The emulator uses true RTL copies from the MOSAIX design for nearly all control and data path logic. Lightweight behavioral models were developed for items like the pixel matrix readout, including sufficient features to verify the backend system. Critically, the emulator incorporates power- and state-aware modeling. For example, it will refuse a command to a tile whose power switch is modeled as "off," forcing the control software to implement and validate the correct, complex operational sequences.

\subsection{Development and Impact}

Development of the emulator and test system began over a year before the MOSAIX tapeout. As the MOSAIX RTL was still under development, this period was characterized by a "moving target" design, which was managed through a semi-automated CI flow to integrate frequent updates to register maps and logic interfaces, converting SystemRDL and RTL to FPGA-friendly format with additional debug features. Because of this automation, the emulator could provide a working version to the users that matched the latest documentation without significant delays and the development period converged around the time of the MOSAIX tapeout.

The investment in the emulator yielded substantial benefits. It served as a real-time application environment for developing and verifying the entire test infrastructure (hardware, firmware, and software). It uncovered operational complexities of the MOSAIX that the software has to manage, that would have caused months of delays if discovered only after receiving the silicon, a fact underscored by the significant time required to develop the software to manage them. For example, this was observed in the dependency of a fully working and configured service layer of the MOSAIX before one can even power on the global digital supply and communicate with the tiles.
Furthermore, the emulator has proven to be an invaluable training tool, allowing more than 50 users to gain hands-on experience with the chip months before its arrival.

\section{Status and Conclusion}

The foundational infrastructure of the qualification system for the MOSAIX sensor is complete and verified. The hardware, firmware, and core software API have been tested and validated against the FPGA emulator. Primary testing sequences are developed and are running on the full setup using the emulator. Work is now focused on implementing advanced test sequences.

The key lesson from this work is the value of early system integration enabled by a high-fidelity emulator. The scale of the software development and the complexity of the chip's operational procedures were significant challenges that have been effectively mitigated by this approach, by providing more time for this development and debugging ahead of the delivery of the MOSAIX.

With the first MOSAIX wafers expected during the beginning of 2026, the qualification system is ready to begin the characterization campaign immediately. This readiness is crucial to providing the necessary design feedback by mid-2026, keeping the ALICE ITS3 upgrade on schedule.

\flushbottom
\newpage
\clearpage




\end{document}